\documentclass[aps,pra,epsfig,twocolumn,showpacs]{revtex4}

\def\be{\begin{equation}}
\def\ee{\end{equation}}
\def\bea{\begin{eqnarray}}
\def\eea{\end{eqnarray}}
\def\bma{\begin{mathletters}}
\def\ema{\end{mathletters}}
\def\bi{\begin{itemize}}
\def\ei{\end{itemize}}

\def\C{\hbox{$\mit I$\kern-.7em$\mit C$}}

\begin{document}

\title{Local cloning of Bell states and distillable entanglement}

\author{Sibasish Ghosh}
\email{sibasish@imsc.res.in}
\affiliation{Institute of Mathematical Sciences, C. I. T. Campus, Taramani, Chennai 600113, India}

\author{Guruprasad Kar}
\email{gkar@imsc.res.in}
\affiliation{Institute of Mathematical Sciences, C. I. T. Campus, Taramani, Chennai 600113, India}

\author{Anirban Roy}
\email{anirb@imsc.res.in}
\affiliation{Institute of Mathematical Sciences, C. I. T. Campus, Taramani, Chennai 600113, India}

\date{\today}

\begin{abstract}
The necessary and sufficient amount of entanglement required for cloning of orthogonal Bell states by local operation and classical communication is derived, and using this result, we provide here some additional examples of reversible, as well as irreversible states.
\end{abstract}


\maketitle

\section{Introduction}
  
Nonorthogonal states cannot be cloned exactly \cite{woo82}, whereas orthogonal states can be. But multipartite orthogonal states can not be cloned only by local operation and classical communication (LOCC) in general without using entangled ancilla state, as otherwise that would imply creation of entanglement by LOCC. In this scenario, if one allows these far apart parties to share some known entangled states as extra resource in the form of blank copy state and machine states (we together call them as {\it ancilla}), cloning of orthogonal multipartite (entangled in general) will be possible. 

Through out this paper, we shall consider the four Bell states as:
\begin{equation}
\label{bell}
\begin{array}{lcl}
|B_1\rangle &=& \frac{1}{\sqrt{2}}(|00\rangle + |11\rangle ),\\
|B_2\rangle &=& \frac{1}{\sqrt{2}}(|00\rangle - |11\rangle ),\\
|B_3\rangle &=& \frac{1}{\sqrt{2}}(|01\rangle + |10\rangle ),\\
|B_4\rangle &=& \frac{1}{\sqrt{2}}(|01\rangle - |10\rangle ),
\end{array}
\end{equation}
where one of the particles ({\it i.e.}, one of the qubits) is held by Alice and the other is held by Bob. 

We show in section II that any two Bell states can be cloned by sharing just 1 ebit free entanglement in the ancilla state. Cloning of four Bell states is discussed in section III. This has been generalized, in section IV, for $1 \rightarrow N$ copy cloning of Bell states. In section V, we provide some new examples of quasi-pure states \cite{quasipure}. We provide some new examples of irreversible states, in section VI. We summarize our results in section VII.

\section{$1 \rightarrow 2$ exact cloning of two Bell states}

Now consider the simplest case where the state may be any two of $|B_i\rangle$'s. If they share, as ancilla, two known ebits, they can first locally distinguish the above-mentioned two Bell states, and then they can locally transform the two shared ebit states to two copies of the corresponding Bell states. In fact, one can have the same result by one ebit of shared ancilla state (although the cloned state will remain unknown). How to achieve this? Let Alice and Bob share either $|B_1\rangle$ or $|B_3\rangle$ (see equation (\ref{bell})). If they share (as ancilla) the known state $|B_1\rangle$, then by applying C-NOT operation locally (where the qubit of unknown Bell state is the source and the qubit of the ancilla will be the target, for both Alice and Bob), they will share two copies of the unknown Bell state. The same result can be obtained (modulo some overall phases) for any two bell states from the set given in (\ref{bell}), by, first local unitarily transforming (or, expressing the local bases in terms of some other local bases) the two Bell states to the states $|B_1\rangle$, $|B_3\rangle$ (expressed in the new basis), then making the local cloning of these later two Bell states (using 1 ebit of entanglement), and finally local unitarily transforming (or, expressing the local bases in terms of some other local bases) $|B_1\rangle^{\otimes 2}$ and $|B_3\rangle^{\otimes 2}$ to the respective two copies of the initially given set of two Bell states.    

It is to be noted here that one ebit of free ({\it i.e.}, distillable) entanglement is necessary in the shared ancilla state. If not, let $\rho$ be any shared bipartite ancilla state (where the distillable entanglement of $\rho$ is less than 1 ebit), for which, under LOCC, any two Bell states $|B_1\rangle$ and $|B_2\rangle$ (say) can be exactly cloned. This immediately shows that, the initially shared separable state ${\rho}_{sep} = \frac{1}{2} (P[|B_1\rangle] + P[|B_2\rangle])$ of Alice and Bob will be transformed (together with the above-mentioned shared ancilla state $\rho$) as $${\rho}_{sep} \otimes \rho \rightarrow \frac{1}{2}(P[|B_1\rangle \otimes |B_1\rangle] + P[|B_2\rangle \otimes |B_2\rangle]),$$
due to linearity of the $1 \rightarrow 2$ cloning operation \cite{projector}. The final state has 1 ebit of distillable entanglement (across the Alice : Bob cut) \cite{ghosh01}, while total initial state has less than 1 ebit of distillable entanglement -- a contradiction. So, whatever shared ancilla state $\rho$ we take (for $1 \rightarrow 2$ cloning of two Bell states), it must have at least one ebit of distillable entanglement.

Above-mentioned argument can be directly generalized to show that $(N - 1)$ ebits of free entanglement ({\it i.e.}, distillable entanglement) in the shared ancilla state is necessary as well sufficient to have $1 \rightarrow N$ exact cloning of any given set of two Bell states (given in (\ref{bell})), by LOCC only \cite{note0}. This result immediately shows that the distillable entanglement of ${\rho}_N^{(2)}$ where ${\rho}_N^{(2)} = \frac{1}{2} (P[|B_1\rangle^{\otimes N}] + P[|B_3\rangle^{\otimes N}])$, in the Alice : Bob cut, is given by $E_D\left({\rho}_N^{(2)}\right) = N - 1$.  

\section{$1 \rightarrow 2$ exact cloning of four Bell states}

Let us now come to the case where Alice and Bob has to clone one (but unknown) of the four Bell states, given in (\ref{bell}). It can be easily shown that if Alice and Bob share three ebits ({\it i.e.}, three known maximally entangled states of two qubits),  they can prepare two copies of that unknown state, locally. Because with one shared ebit, Alice can teleport her part (of the unknown Bell state) to Bob, and Bob can distinguish these four Bell states, and hence they can transform locally the rest two copies of the Bell states to the desired Bell states. 

We now show that for cloning any one of the four Bell states, given in (\ref{bell}), locally, two ebits of shared ancilla state is necessary. Consider the state where two far apart parties Alice and Bob are sharing two copies of one of the four Bell states with equal probabilities. Thus the shared state is:
\begin{equation}
\label{smolinstate}
{\rho}_S = \frac{1}{4} \sum_{i = 1}^{4} P\left[|B_i\rangle_{A_1B_1} \otimes |B_i\rangle_{A_2B_2}\right],
\end{equation}
where Alice is holding the qubits $A_1$, $A_2$, while Bob is holding the qubits $B_1$, $B_2$. This state is interestingly separable in the $A_1A_2 : B_1B_2$ cut \cite{smolin00}. If cloning of the four Bell states is possible with shared ancilla state $\rho$, by using LOCC only, we then have    
$${\rho}_S \otimes \rho \rightarrow \frac{1}{4} \sum_{i = 1}^{4} P\left[|B_i\rangle \otimes |B_i\rangle \otimes |B_i\rangle\right].$$
The final state has 2 ebits of distillable entanglement \cite{yang03}, and hence $\rho$ must have at least 2 ebits of distillable entanglement.

Now we are going to show that two ebits of distillable entanglement, in the shared ancilla state, is also sufficient for exact cloning of the four Bell states. In order to do this, let us first consider teleportation of two-qubit states (locally) via the state ${\rho}_S$, given in equation (\ref{smolinstate}). Interestingly, this state can also be written as
$${\rho}_S = \frac{1}{4} \sum_{i = 1}^{4} P\left[|B_i\rangle_{A_1A_2} \otimes |B_i\rangle_{B_1B_2}\right],$$
where the four qubits $A_1, A_2, B_1, B_2$ might be far apart, where Alice(i) is holding the qubit $A_i$, and similarly for Bob. Now we consider the teleportation of a two-qubit state shared between Alice(1) and Bob(1). Let Alice(1) and Bob(1) are teleporting their qubits to Alice(2) and Bob(2) respectively, using the same teleportation protocol, namely the standard teleportation protocol of Bennett et al. \cite{benn93}, for exactly teleporting an unknown qubit through the channel state $|B_1\rangle$. This teleportation ({\it i.e.}, teleportation of any two-qubit state from Alice(1) : Bob(1) cut to Alice(2) : Bob(2) cut, via the shared channel state ${\rho}_S$, and using the exact protocol of Bennett et al. \cite{benn93} for the channel state $|B_1\rangle$, for each qubit) can be represented in terms of the following maps (see \cite{ghoshpra01}):
\begin{equation}
\label{teleportationmap}
{\sigma}_i^{A_1} \otimes {\sigma}_j^{B_1} \rightarrow  {\delta}_{ij} {\sigma}_i^{A_2} \otimes {\sigma}_j^{B_2},
\end{equation}
where $i, j = 0, 1, 2, 3$, and where ${\sigma}_0^{A_1}$ is the identity operator acting on the Hilbert space of the qubit $A_1$, ${\sigma}_1^{A_1}$ is the Pauli matrix ${\sigma}_x$, acting on the Hilbert space of the qubit $A_1$, etc., etc. (see ref. \cite{mapexplanation} for an explanation of the map (\ref{teleportationmap})). Using this map, one can easily see that any Bell state (from equation (\ref{bell})), shared between Alice(1) and Bob(1), can be exactly teleported to Alice(2) and Bob(2) \cite{note1}. 

Consider now the following state of six qubits $A_i$, $B_i$ ($i = 1, 2, 3$):
\begin{equation}
\label{rho3}
{\rho}^{(3)} = \frac{1}{4} \sum_{i = 1}^{4} P\left[|B_i\rangle_{A_1B_1} \otimes |B_i\rangle_{A_2B_2} \otimes |B_i\rangle_{A_3B_3}\right].
\end{equation}
This state is a quasi-pure state \cite{chen03, werner03}. It has 2 ebits of distillable entanglement in the $A_1A_2A_3 : B_1B_2B_3$ cut \cite{yang03}, while it can also be created using 2 ebits of entanglement. Here we are going to provide a simple preparation procedure of this state, starting from 2 ebits of free entanglement.  
Let Alice and Bob start with the following shared 2 ebit state $D = P[|B_1\rangle_{A_1B_1} \otimes |B_1\rangle_{A_2B_2}] \otimes \frac{1}{2}(P[|B_1\rangle_{A_3B_3}] + P[|B_2\rangle_{A_3B_3}])$. Now, with equal probability, Bob applies either nothing or he applies the Pauli matrix ${\sigma}_x$ on each of his three qubits. Next Alice and Bob both apply locally two consecutive C-NOT operations on their respective qubits, taking the third pair of qubits as target (for both of these two C-NOT operations), while the first two pairs of qubits will act as source. This will directly give rise to the state ${\rho}^{(3)}$ of equation (\ref{rho3}). 

The two reduced density matrices of ${\rho}^{(3)}$, corresponding to the subsets $\{A_1, B_1, A_2, B_2\}$ and  $\{A_1, B_1, A_3, B_3\}$ are the same Smolin state ${\rho}_S$. Thus if Alice(1) and Bob(1) jointly teleport (using the map given in (\ref{teleportationmap})) the unknown Bell state, it will be reproduced both between Alice(2), Bob(2) as well as between Alice(3), Bob(3). Thus two ebits of shared ancilla state (shared between Alice's side ($A_1, A_2, A_3$) and Bob's side ($B_1, B_2, B_3$)) is sufficient to have the $1 \rightarrow 2$ cloning of one of the four Bell states, given in (\ref{bell}).    

\section{$1 \rightarrow N$ exact cloning of four Bell states}

Interestingly $1 \rightarrow 3$  cloning of unknown Bell state, taken from the set given in (\ref{bell}), by LOCC, also requires 2 ebits of entanglement.  If Alice and Bob share the following 8-qubit ancilla state  $\rho^{(4)} = \frac{1}{4} {\sum}_{i = 1}^{4} P[|B_i\rangle_{A_1B_1} \otimes |B_i\rangle_{A_2B_2} \otimes |B_i\rangle_{A_3B_3} \otimes |B_i\rangle_{A_4B_4}]$,  using the same procedure as above, they can have the desired  $1 \rightarrow 3$  cloning of the unknown Bell state. Interestingly, single copy of this 8-qubit state can also be prepared from 2 ebits of free entanglement. We now describe a simple process of this state. Alice and Bob first start with the following shared 2 ebit state:  $D =  \frac{1}{2}(P[|B_1\rangle_{A_1B_1}] + P[|B_3\rangle_{A_1B_1}]) \otimes  P[|B_1\rangle_{A_2B_2} \otimes |B_1\rangle_{A_3B_3}] \otimes \frac{1}{2}(P[|B_1\rangle_{A_4B_4}] + P[|B_2\rangle_{A_4B_4}])$. Alice and Bob both now apply two consecutive C-NOT operations, where the first pair of qubits are taken as source, while the second and third pairs of qubits are targets.  In the next step, Alice and Bob both apply locally three consecutive C-NOT operations on their respective qubits, taking the fourth pair of qubits as target (for both of these two C-NOT operations), while the first three pairs of qubits will act as source. This will directly give rise to the state ${\rho}^{(4)}$. 

This result can be easily generalized for $1 \rightarrow N$ cloning of four Bell states, using the shared ancilla state \\ \\
$\rho^{(N+1)} = $ 
\begin{equation}
\label{rhoN}
\frac{1}{4} \sum_{i = 1}^{4} P[|B_i\rangle_{A_1B_1} \otimes |B_i\rangle_{A_2B_2} \ldots \otimes |B_i\rangle_{A_{(N+1)}B_{(N+1)}}].
\end{equation}
This ancilla state $\rho^{(N+1)}$ has N ebits of distillable entanglement when $N$ is even, and $N-1$ for $N$ odd. But $\rho^{(N+1)}$ being reversible \cite{chen03}, $\rho^{(N+1)}$ can be prepared using $N$ and $N-1$ ebits for even and odd cases respectively.   

\section{Examples of quasi-pure states}

Using these above results of cloning we now provide some new examples of quasi-pure states, each of which is reversible. Consider the following $2N$ party entangled state (where $N$ is odd) 
$$\rho(p) = {\sum}_{i=1}^4 p_i P[|B_i\rangle^{\otimes N}],$$ 
where, for all the $N$ copies of the Bell state $|B_i\rangle$, one qubit is with Alice, while the other qubit is with Bob, and where $0 \le p_i \le \frac{1}{2}$ for all $i = 1, 2, 3, 4$. This state can be prepared by the following process. Consider the following density matrix $\rho_B = {\sum}_{i=1}^4 p_i P[|B_i\rangle]$ with the same $p_i$ as above ($\rho_B$ is obviously separable). This state can be prepared locally by Alice and Bob. Alice and Bob also now locally prepare the state $\rho^{(N + 1)}$, given as in equation (\ref{rhoN}), using $(N - 1)$ ebits of free entanglement \cite{chen03}. Take $\rho_B$ with $\rho^{(N + 1)}$  and apply the previous $1 \rightarrow N$ cloning operation, the final state will become $\rho(p)$. So the entanglement cost \cite{benn96, hay01} $E_c$ of the state $\rho(p)$ will be at most $(N - 1)$. But now Alice and Bob locally apply $(N - 1)$ consecutive C-NOT operations on their respective sets of qubits (where the $N$-th qubit pair is taken as the target and each of the first $(N - 1)$ qubit pairs are taken as source), and then they locally distinguish the two sets of Bell states $\{|B_1\rangle, |B_2\rangle\}$, $\{|B_3\rangle, |B_4\rangle\}$ in the target qubit pairs. This will then directly give rise to $(N - 1)$ copies of $|B_1\rangle$ or $|B_3\rangle$ (depending upon the measurement outcomes), and hence, $(N - 1)$ ebits of entanglement can be distilled from ${\rho}(p)$ (see also \cite{chen03}).
Hence, for this state, entanglement cost and distillable entanglement are same (and its value is equal to $(N - 1)$), providing an example of reversibility. 

\section{Examples of irreversible states}

Now we generalize the example of irreversibility in \cite{vidal02}, to the case where $N$ no. of Bell states are involved. It has been shown in \cite{vidal02} that for the state
$${\sigma}_1 = p P[|B_1\rangle] + (1 - p) P[|B_2\rangle],$$
where $0 < p < 1$ and $p \ne \frac{1}{2}$, we have  $E_c({\sigma}_1) = $ \\ 
$H_2[\frac{1}{2} + \sqrt{p(1-p)}]$. But its distillable entanglement is given by $E_D({\sigma}_1) =  1 - H_2(p)$,  where for $0 \le x \le 1$, $H_2(x) = -x {\rm log}_2 x - (1 - x) {\rm log}_2 (1 - x)$ \cite{benn96}. Thus we have here
\begin{equation}
\label{sigma1}
E_c({\sigma}_1) > E_D({\sigma}_1).
\end{equation}
Next we consider the state 
$${\sigma}_N = p P[|B_1\rangle^{\otimes N}] + (1 - p) P[|B_2\rangle^{\otimes N}],$$
where $0 < p < 1$ and $p \ne \frac{1}{2}$. And we are going to show now that  $E_c({\sigma}_N) > E_D({\sigma}_N)$.  First of all, we show that ${\sigma}_N$ can be prepared locally from the state ${\sigma}_1$ together with $(N - 1)$ ebits of free entanglement $|B_1\rangle^{\otimes (N - 1)}$. In this direction, first both Alice and Bob locally apply the Hadamard transformation on the respective qubits of the state ${\sigma}_1$. Thus we will have the following transformation
$${\sigma}_1 \otimes P[|B_1\rangle^{\otimes (N - 1)}] \rightarrow$$
$$    (p P[|B_1\rangle] + (1 - p) P[|B_3\rangle]) \otimes  P[|B_1\rangle^{\otimes (N - 1)}].$$   
Now both Alice and Bob use C-NOT operations locally and sequentially, where the source qubit belongs to the state ${\sigma}_1$, while the target qubits will come from each of the $(N - 1)$ Bell states $|B_1\rangle$ (see the case of cloning of the two Bell states $|B_1\rangle$ and $|B_3\rangle$). Thus we have now the following transformation:
$$(p P[|B_1\rangle] + (1 - p) P[|B_3\rangle]) \otimes  P[|B_1\rangle^{\otimes (N - 1)}] \rightarrow$$
$$   p P[|B_1\rangle^{\otimes N}] + (1 - p) P[|B_3\rangle^{\otimes N}]$$ 
$$\equiv {\sigma}^{\prime}_N~~ ({\rm say}).$$
Again Alice and Bob will locally apply Hadamard transformations on each of the $N$ qubits of their sides, so finally, we will have the desired state ${\sigma}_N$.  As the transformation ${\sigma}_1 \otimes P[|B_1\rangle^{\otimes (N - 1)}] \rightarrow  {\sigma}_N$ has been achieved above by using local unitary transformation $U_A \otimes V_B$ in the Alice : Bob cut, therefore, $E_c({\sigma}_N) = E_c({\sigma}_1 \otimes P[|B_1\rangle^{\otimes (N - 1)}])$ (in the Alice : Bob cut). And, as from equation (17) of ref. \cite{vidal01}, it is known that  $E_c({\sigma}_1 \otimes P[|B_1\rangle^{\otimes (N - 1)}]) = E_c({\sigma}_1) + (N - 1)$, therefore, 
\begin{equation}
\label{ecsigman}
E_c({\sigma}_N) =  E_c({\sigma}_1) + (N - 1) = H_2[\frac{1}{2} \sqrt{p(1 - p)}] + (N - 1).
\end{equation}
 On the other hand, using the reverse local unitary operation $U_A^{\dagger} \otimes V_B^{\dagger}$, we have the following transformation:
$${\sigma}_N \rightarrow {\sigma}_1 \otimes P[|B_1\rangle^{\otimes (N - 1)}].$$
And so, $E_D({\sigma}_N) = E_D({\sigma}_1 \otimes P[|B_1\rangle^{\otimes (N - 1)}])$. Now considering the invariance under local unitary operations and  sub-additivity property of relative entropy of entanglement $E_R(.)$, we have $E_R({\sigma}_N) = E_R({\sigma}_1 \otimes P[|B_1\rangle^{\otimes (N - 1)}]) \le E_R({\sigma}_1) + (N - 1) = 1 - H_2(p) + (N - 1)$ \cite{vedral97}. And, as $E_R(.)$ is an upper bound of $E_D(.)$ \cite{rains99a}, therefore, $E_D({\sigma}_N) \le E_R({\sigma}_N) \le E_R({\sigma}_1) + (N - 1) = E_D({\sigma}_1) + (N - 1)$ (as, for the state ${\sigma}_1$, we have $E_D({\sigma}_1)  =  1 - H_2(p)$ \cite{benn96}, and also $E_R({\sigma}_1) = 1 - H_2(p)$ \cite{vedral97}). But as we can distill $E_D({\sigma}_1)$ ebit of entanglement from a single copy of ${\sigma}_1$, therefore, from the state ${\sigma}_1 \otimes P[|B_1\rangle^{\otimes (N - 1)}]$, we can distill at least  $E_D({\sigma}_1) + (N - 1)$  ebits of entanglement. And so, 
\begin{equation}
\label{sigmaned}
E_D({\sigma}_N) =  E_D({\sigma}_1) + (N - 1) = N - H_2(p).
\end{equation}
Using equations (\ref{sigma1}), (\ref{ecsigman}) and (\ref{sigmaned}), we have  $E_c({\sigma}_N) > E_D({\sigma}_N)$. Hence the state ${\sigma}_N$ is irreversible.
      
\section{Conclusion}

In summary, we have shown that the minimum resource (in terms of free entanglement) required to make $1 \rightarrow N$ copy cloning by LOCC of unknown Bell states, from any known set of two Bell states, is $(N - 1)$. On the other hand, it has been shown here that the minimum resource required to make $1 \rightarrow N$ copy cloning by LOCC of unknown Bell states, from the set of four Bell states, is different for odd and even cases. In even case it is just equal to $N$ ebit whereas in odd case it is $(N - 1)$. We have also given here new examples of reversible states as well as irreversible states, formed by using multiple copies of Bell states.

Our results show that in order to have $1 \rightarrow N$ exact cloning of any given set of three Bell states, it is necessary to have at least $(N - 1)$ ebits of free entanglement in the shared ancilla state, while $N$ ebits of free entanglement (in the shared ancilla state) is sufficient in this case, if $N$ is even, and $(N - 1)$ ebits of free entanglement (in the shared ancilla state) is sufficient in this case, if $N$ is odd. Existence of more tight bound on the sufficient condition, in this case, is an open question.\\ \\              

{\noindent {\bf Acknowledgement :} The authors are thankful to Debasis Sarkar for useful discussions, in particular, about teleportation via the Smolin state.}


\begin{thebibliography}{00}
\bibitem{quasipure} A bipartite state ${\rho}_{AB}$, shared by two far apart parties Alice and Bob, is said to be {\it quasi-pure} if it can be reversibly transformed, by LOCC, to the state $\sum_{i = 1}^{m} p_i {\rho}^{(i)}_{A_1B_1} \otimes \left|{\phi}_i\right\rangle_{A_2B_2}\left\langle{\phi}_i\right|$, where ${\rho}^{(i)}_{A_1B_1}$'s are pairwise orthogonal separable states, which can be exactly distinguished by LOCC, and where $A_1$, $A_2$ are subsystems of the system $A$, while $B_1$, $B_2$ are subsystems of the system $B$.
\bibitem{projector} For any (normalized) pure state $|\psi\rangle$, $P[|\psi\rangle]$ is the projector $| \psi \rangle \langle \psi |$ of $|\psi\rangle$.
\bibitem{note0} This can be achieved in the following way. At first, we assume that the two unknown Bell states, which have to be cloned to get $N$ copies, are $|B_1\rangle_{A_1B_1}$ and $|B_3\rangle_{A_1B_1}$. Also let Alice and Bob share the $(N - 1)$-ebit ancilla state $|B_1\rangle_{A_2B_2} \otimes |B_1\rangle_{A_3B_3} \otimes \ldots \otimes |B_1\rangle_{A_NB_N}$, where the $N$ qubits $A_1$, $A_2$, $\ldots$, $A_N$ are with Alice and the $N$ qubits $B_1$, $B_2$, $\ldots$, $B_N$ are with Bob. Both Alice and Bob first locally apply the C-NOT operations, taking $A_1$ as source, $A_2$ as target (for Alice's C-NOT), and  taking $B_1$ as source, $B_2$ as target (for Bob's C-NOT). After that, both Alice and Bob locally apply C-NOT operations, taking $A_2$ as source, $A_3$ as target (for Alice's C-NOT), and  taking $B_2$ as source, $B_3$ as target (for Bob's C-NOT). Continuing this process, they will finally come up with $N$ copies of the unknown Bell states. For given any two Bell states from (\ref{bell}), in order to have their $1 \rightarrow N$ cloning by LOCC, we shall apply the same procedure (as has been done in the case of $1 \rightarrow 2$ cloning of any two Bell states) of firstly transforming them local unitarily to $|B_1\rangle$, $|B_3\rangle$, and then we will have to apply the above-mentioned argument.
\bibitem{note1} One may be surprised that although there is no entanglement in the (Alice(1) Bob(1)) : (Alice(2) Bob(2)) cut of the state ${\rho}_S$ (given in (\ref{smolinstate}), each of the four Bell states is exactly getting teleported from the side of Alice(1), Bob(1) to the side of Alice(2), Bob(2). First of all, no entanglement is being created by this teleportation process, which is here a LOCC among the four parties Alice(1), Alice(2), Bob(1) and Bob(2). Secondly, just four pairwise orthogonal states $|B_1\rangle$, $|B_2\rangle$, $|B_3\rangle$, $|B_4\rangle$, (given in (\ref{bell})) from infinitely many states (forming the four dimensional Hilbert space of two qubits), are being exactly teleported in this way -- no other (pure) state can be exactly teleported. And we know that this phenomenon holds good even when Alice(1) Bob(1) share a product state with Alice(2) Bob(2).
\end{thebibliography}
\end{document}